\documentclass[twocolumn,twoside,slac_two]{revtex4}
\usepackage{graphicx}
\usepackage{fancyhdr}
\usepackage{hyperref}
\hypersetup{
    colorlinks=true,
    urlcolor=magenta,
}
\begin{document}

\title{Application of a full chain analysis using neutron monitor data for space weather studies}


\author{A. Mishev}
\affiliation{Space Climate Research Unit, University of Oulu, Oulu, Finland}
\author{I. Usoskin}
\affiliation{Space Climate Research Unit, University of Oulu, Oulu, Finland}
\affiliation{Sodankyl\"a Geophysical Observatory (Oulu unit), University of Oulu, Oulu, Finland}

\begin{abstract}
An important topic in the field of space weather is the precise assessment of the contribution of galactic cosmic rays and solar energetic particles on air crew exposure, specifically during eruptive events on the Sun. Here we present a model, a full chain analysis based on ground based measurements of cosmic rays with neutron monitors, subsequent derivation of particle spectral and angular characteristics and computation of dose rate. The model uses method for ground level enhancement analysis and newly numerically computed yield functions for conversion of secondary particle fluence to effective dose and/or the ambient dose equivalent. The precise an adequate information about the solar energetic particle spectra (SEPs) is the basis of the model. Since SEPs  possess an essential isotropic part, specifically during the event onset, the angular characteristics should be also derived with good precision. This can be achieved using neutron monitor data during a special class of  SEP events – the ground level enhancements (GLEs). A precise analysis of SEP spectral and angular characteristics using neutron monitor (NM) data requires realistic modelling of propagation of those particles in the Earth's magnetosphere and atmosphere. On the basis of the method representing a sequence of consecutive steps, namely a detailed computation of the SEP asymptotic cones of acceptance, NM rigidity cut-off and application of convenient optimization procedure, we derive the rigidity spectra and anisotropy characteristics of GLE particles.  For the computation we use newly computed yield function of the standard sea-level 6NM64 neutron monitor  for primary proton and alpha CR nuclei as well as 6NM64  yield function at  altitudes ranging from the sea level up to 5000 m above the sea level. We derive the SEP spectra and pitch angle distributions in their dynamical development throughout the event.  Subsequently on the basis of the derived spectra and angular characteristics and previously computed yield functions we calculate the effective dose and/or ambient dose equivalent during the GLE. Several examples are shown.  The derived results are compared with the previously obtained assessments and are briefly discussed. 
\end{abstract}

\maketitle

\thispagestyle{fancy}


\section{INTRODUCTION}
The Earth's geo-aerospace is continuously bombarded by high, very high and ultra high energy charged particles, which penetrate deep within the atmosphere. They are known as cosmic rays (CRs), consisting mostly of protons and $\alpha-$particles with small fraction of heavier nuclei \cite{Gaisser2010}. The majority of CRs originate from the Galaxy, called galactic cosmic rays (GCRs) with intensity varying by 10--20$\%$ modulated by solar wind and depending on the level of the solar activity. Therefore, the flux of primary CRs follows the 11-year solar cycle and responds to solar wind variations and transient phenomena \cite{Dorman04}. The Earth is also hit sporadically by high intensity, but with low occurrence rate solar energetic particles (SEPs), accelerated during explosive energy releases on the Sun, which may sometimes produce an atmospheric cascade, leading to a ground level enhancement (GLE) \cite{Dorman04}. Primary CR particles penetrate into the Earth's atmosphere and produce complicated nuclear-electromagnetic-muon cascade, which consists of large amount of secondary particles. In such a cascade, only a fraction of the initial primary particle energy reach the ground as secondaries, most of the primary particle's energy is released in the atmosphere by ionization \cite{Bazilevskaya08}. Therefore CR particles of both galactic and solar origin significantly affect the radiation environment and accordingly exposure at typical commercial flight altitudes of about 35 kft (10 668 m above the sea level) \cite{Vainio09}. In fact, GCRs and SEPs are the most significant contributors to the increased compared to the ground level radiation exposure of aircrew, specifically over the polar region, where the magnetospheric shielding is not as strong as at middle and equatorial latitudes. 

According to the common definition, space weather refers to the dynamic, variable conditions on the Sun, solar wind and Earth’s magnetosphere and ionosphere, that can compromise the performance and reliability of spacecraft and ground-based systems and can endanger human life or health. An important topic of space weather is the assessment of aircrew exposure due to CR, specifically during GLEs events \cite{Lilensten2009}. Because  the aircrews are subject to increased exposure, particularly during intercontinental flights over the Pole due to the increased intensity of secondary CRs, the radiation protection of aircrews became a subject of radiation protection regulations, namely Publication 60 of the International Commission on Radiological Protection \cite{ICRP1991},  where the exposure of flying personnel to cosmic radiation is recommended to be regarded as occupational. Accordingly in EU, the radiation protection of aircrew members, has been legally regulated in the EU since 1996 \cite{EURATOM1996} i.e. there is a suggestion of measures to assess the individual doses of both cockpit and cabin crew. Those recommendations are spread on different timescales, from single flights, where exposure would be due to SEPs, to decadal exposure or even whole carrier exposure. 

In order to mitigate those effects a detailed study is necessary, considering explicitly such features as anisotropy of SEPs, their spectral features as well as their time evolution during major solar proton events.  The flux of primary CRs, both of galactic and/or solar origin is influenced by the spatial-temporal variability of the complex magnetospheric and interplanetary conditions. In addition, the secondary particles in the cascade are characterized by large diversity as type (hadrons such protons, neutrons and mesons, electromagnetic particles such as electrons, positrons and gammas, etc…), distributed in a wide energy range i.e. the radiation field is a complex mix of different particles possessing different radiation impact \cite{Ferrari01}. Therefore, the task to estimate the ambient dose equivalent, accordingly the air-crew exposure due to CR of galactic and solar origin is not trivial, since it depends on geographic position, altitude (the distribution as intensity and energy of the secondary particles) and solar activity \cite{Spurny96, Shea2000}. 

Therefore the assessment of radiation dose hazard due to GCR and/or SEPs involves several consecutive steps. In the first step fluxes of GCR and SEPs must be precisely determined outside the magnetosphere. Subsequently their propagation through the atmosphere and magnetosphere of the Earth should be carefully modelled in order to derive a realistic information of the secondary CR flux. Finally the secondary particle flux is converted to effective and/or ambient dose equivalent.
It was shown that the dose rate can be computed as a function of the geomagnetic rigidity cut-off and altitude using a full Monte Carlo simulation of the atmospheric cascade \citep{Ferrari01, Roesler02}.

Nowadays as a result of numerical methods, based on enhanced knowledge of high-energy interactions and nuclear processes, several models aiming to estimate the dose rate (effective and/or ambient dose equivalent) at flight altitudes due to primary CR radiation based on Monte Carlo simulations have been proposed \cite{Ferrari01, Roesler02, Sato08, Matthia2008, Mertens2013, Mishev15a, Mishev15b}. In this paper, we describe a numerical model to estimate effective and/or ambient dose equivalent at flight altitudes. The model uses NM data in order to derive the SEPs spectral and angular features. Subsequently on the basis on newly computed yield functions for conversion of particle fluence to dose and using the derived spectra and angular characteristics as input we compute the effective dose and ambient dose equivalent.

\section{THE MODEL FOR AIRCREW EXPOSURE}
It is known that the dose rate can be computed as a function of the geomagnetic rigidity cut-off (particularly for GCR and GLE particles) and altitude using a full Monte Carlo simulation of the atmospheric cascade and subsequent conversion of secondary particle flux to effective and/or ambient dose equivalent \cite{Pelliccioni2000}. Here we are using a formalism based on pre-computed effective and ambient dose equivalent yield functions derived with extensive Monte Carlo simulation of primary CR particle induced atmospheric cascade. The formalism of pre-computed yield functions is well known, widely used and easy for application. Thus, the yield functions computed once using very high statistics allow one to achieve very high precision, specifically in the high-energy part of the spectrum and to avoid subsequent extensive computations once the yield function is computed. The effective dose and/or the ambient dose equivalent in any particular case are obtained as an integral of the product of the primary particle spectrum and the pre-computed yield functions. 

For the assessment of spectral and angular characteristics of SEPs we are using data from the global NM network (Fig.1) and convenient optimization procedure (see bellow subsection II.A).

\begin{figure} [!ht]
\centering \resizebox{\columnwidth}{!}{\includegraphics{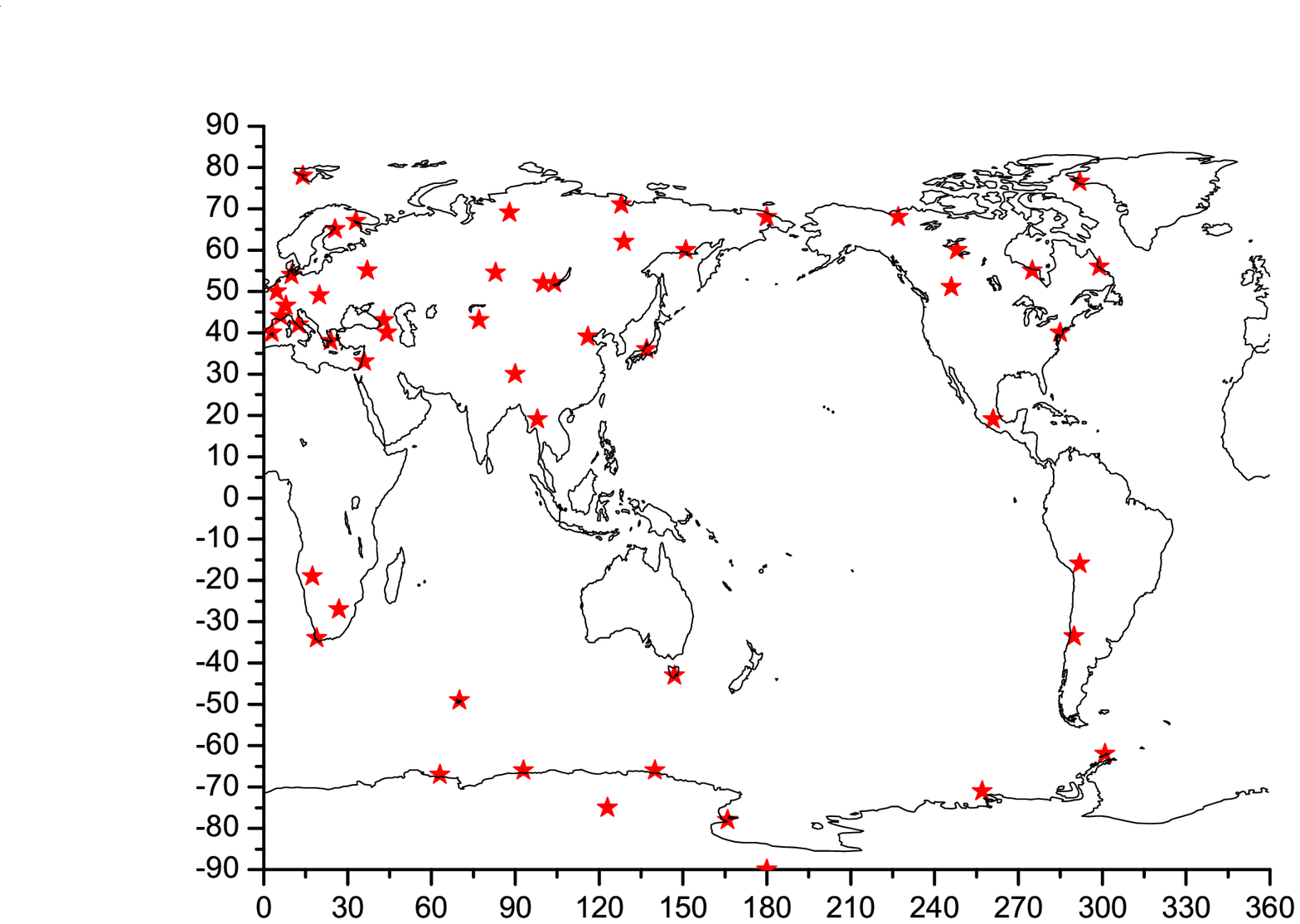}} 
 \caption{Global neutron monitor network (the red stars denote NM stations), experimental data used to derive the SEPs spectra and angular characteristics during GLE events.  \label{fig1}}
 \end{figure}

\subsection{Derivation SEPs spectra and angular characteristics using data from the global NM network}
The spectral and angular characteristics of SEPs using the information retrieved by ground based neutron monitors would be estimated using the relationship between the NM count rates and the primary particle flux via the NM yield function, which considers the full complexity of particle transport in the geomagnetosphere and in the Earth atmosphere and the detector response i.e. registration efficiency and effective area itself. The analysis of a GLEs using NM data consists of several consecutive steps: computation of asymptotic viewing cones and rigidity cut-offs of the NMs by modelling of particle propagation in a model magnetosphere; assumption of an initial guess of the inverse problem; application of an optimization procedure (inverse method) for derivation of the primary SEPs energy spectrum, anisotropy axis direction, pitch angle distribution. The detailed description of the method is given elsewhere \cite{Mishev14a}. The method is similar to that used by \cite{Cramp1997, Vas06}. The relative count rate increase of a given NM is expressed as:

\begin{equation}
\frac{\Delta N(P_{cut})}{N} =\frac{\int_{P_{cut}}^{P_{max}}J_{||sep}(P,t)Y(P)G (\alpha(P,t)) dP}{\int_{P_{cut}}^{\infty}J_{GCR}(P,t)Y(P)dP}
\label{simp_eqn1}
   \end{equation}
\noindent where $J_{||sep}$ is the rigidity spectrum of the primary solar particles in the direction of the maximal flux, $J _{GCR}(P,t)$ is the rigidity spectrum of GCR at given time $t$ with the corresponding modulation, Y(P) is the NM yield function, $G(\alpha(P,t))$ is the pitch angle distribution of SEPs (the pitch angle $\alpha$ is defined as the angle between the asymptotic direction $\vec{\mathcal{P}}$ and the axis of anisotropy $\vec{\mathcal{A}}$ i.e. $\cos(\alpha)$=$\vec{\mathcal{P}}$.$\vec{\mathcal{A}}$, $N$ is the count rate due to GCR, $\Delta N(P_{cut})$ is the count rate increase due to solar particles, $P_{cut}$ is the minimum rigidity cut-off of the station, accordingly $P_{max}$ is the maximum rigidity of SEPs considered in the model, assumed to be 20 GV, which is sufficiently high for SEPs. The fractional increase of the count rate of a NM station represents the ratio between the NM count rates due to SEPs and GCR averaged over 2 hours before the event's onset. Here we use a newly computed NM yield function, which considers the finite lateral extend of cosmic ray induced atmospheric cascade and provides good agreement with experimental latitude surveys as well as other measurements \cite{Mishev13a, Gil15}. In addition in order to reduce and eliminate uncertainties as the application of two attenuation lengths method i.e. normalization of high altitude NM count rates to the sea level during GLE analysis, which includes additional uncertainty, we use NM yield functions computed at various altitudes. This allows one to model the majority of NMs in the network realistically, i.e. the response of each NM is modelled with his own yield function corresponding to the altitude above the sea level. 

Here, the computation of NM rigidity cut-off and asymptotic directions is performed with the MAGNETOCOSMICS code \cite{Desorgher05} using the combination of IGRF geomagnetic model (epoch 2010) as the internal field model and the Tsyganenko 89 model as external field \cite{Tsyganenko89}. This combination provide perfect balance between simplicity and realism \cite{Kudela04, Nevalainen05}. For the galactic cosmic ray (GCR) spectrum we apply a parametrisation based on the force-field model \cite{Gle68, Cab04} with a solar modulation parameter according to \cite{Usoskin11a}. The solution of the inverse problem is performed using the Levenberg-Marquardt method \cite{Lev44, Mar63}, where the optimization is performed by minimization of the difference between the modelled NM responses and the measured NM responses i.e.optimization of the functional $\mathcal{F}$ over the vector of unknowns and $m$ NM stations:

\begin{equation}
\mathcal{F}=\sum_{i=1}^{m} \left[\left(\frac{\Delta N_{i}}{N_{i}}\right)_{mod.}-\left(\frac{\Delta N_{i}}{N_{i}}\right)_{meas.}\right]^{2}
\label{simp_eqn2}
   \end{equation}

\subsection{Computation of effective dose and/or ambient dose equivalent at flight altitudes}
It is known that the effective dose, usually used for radiation protection purposes is not a measurable, therefore, the International commission of radiation protection ICRP suggest the ambient dose equivalent denoted as $H^{*}(d)$, which is the dose equivalent that would be produced by the corresponding expanded and aligned field at a depth $d$ in a International Commission on Radiation Units and Measurements (ICRU) sphere (a sphere with diameter of 30 cm made of tissue equivalent material with a density of 1 $g.cm^{3}$ and a mass composition of 76.2 $\%$ Oxygen, 11.1 $\%$ Carbon, 10.1 $\%$ Hydrogen and 2.6 $\%$ Nitrogen) on the radius vector opposing the direction of the aligned field. The unit for both effective dose and ambient dose equivalent is Sv. The ambient dose equivalent at a depth of d=10mm $H^{*}(10)$  is recommended as a reasonable proxy for the effective dose \cite{Pelliccioni2000}. 

The effective dose rate at a given atmospheric depth $h$ induced by a primary CR particle is given by:

\begin{equation}
E(h, \lambda_{m},\theta,\varphi)= \sum_{i} \int\limits_{T^{'}(\lambda_{m})}^{\infty} \int\limits_{\Omega} J_{i}(T{'}) Y_{i}(T^{'},h) d\Omega dT^{'}
\label{simp_eqn3}
   \end{equation}

\noindent where $T^{'}$ is the energy of the primary CR particle arriving from zenith angle $\theta$ and azimuth angle $\varphi$, $J_{i}(T{'})$ is the differential
energy spectrum of the primary CR at the top of the atmosphere for $i$ component (proton  and/or $\alpha-$particle), $\lambda_{m}$
 is the geomagnetic latitude, $\Omega(\theta,\varphi)$ is a solid angle and $Y_{i}$ is the effective dose yield function.
The corresponding effective dose yield function  $Y_{i}$ is defined as 

\begin{equation}
Y_{i}(T^{'},h) = \sum_{j} \int\limits_{T^{*}}  F_{i,j}(h,T^{'},T^{*}, \theta,\varphi) C_{j}(T^{*}) dT^{*}
\label{simp_eqn4}
   \end{equation}

\noindent where $C_{j}(T^{*})$ is the fluence to effective dose conversion coefficient for a secondary particle of type $j$ (neutron, proton,
 $\gamma$, $e^{-}$, $e^{+}$, $\mu^{-}$, $\mu^{+}$, $\pi^{-}$, $\pi^{+}$) with energy $T^{*}$, $F_{i,j}(h,T^{'},T^{*}, \theta,\varphi)$
 is the fluence of secondary particle of type $j$, produced by a primary particle of type $i$ (proton  and/or $\alpha-$particle)
 with a given primary energy $T^{'}$. The conversion coefficients $C_{j}(T^{*})$ for a particle of type $j$ are obtained using Monte Carlo simulations \cite{Pelliccioni2000, Petoussi2010}. 
 
Accordingly the secondary particle flux produced by primary protons and $\alpha-$particles in a wide energy range is obtained on the basis of extensive high statistics atmospheric cascade simulations using PLANETOCOSMICS code \cite{Desorgher05} assuming a realistic atmospheric model NRLMSISE00 \cite{Picone02}. The yield function considers the complexity of the atmospheric cascade development, since  it brings information of particle fluence and spectrum at a given altitude in the atmosphere, considering the secondary particle attenuation. The newly computed effective dose yield function for 35 kft (typical commercial flight altitude $\approx$ 10.5 km) is shown in Fig.2, within the separate contribution of protons and $\alpha-$particles. The integration in  ~Eq. (\ref{simp_eqn3}) is over the kinetic energy above $T^{'}(\lambda_{m})$=$E_{cut}(R_{c})$, which is defined by the local rigidity cut-off $R_{c}$ for a nuclei of type $i$ at a given geographic location by the expression:

\begin{equation}
 E_{cut,i}=\sqrt{ \left( \frac {Z_{i}} {A_{i}}\right)^{2} R_{c}^{2}+ E_{0}^{2}} - E_{0}
        \label{simp_eqn5}
   \end{equation}
  
\noindent where $E_{0}$ =  0.938 GeV/n is the proton's rest mass.

 \begin{figure} [!ht]
\centering \resizebox{\columnwidth}{!}{\includegraphics{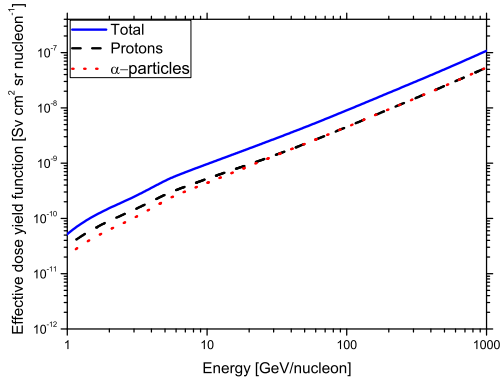}}
 \caption{Effective dose yield function as a function of the energy per nucleon for primary CR protons and $\alpha-$particles as denoted in the legend at 35 kft (altitude of $\approx$ 10.5 km)\label{fig2}}
 \end{figure}

Accordingly, the ambient dose equivalent $H^{*}(10)$ i.e. the dose equivalent produced at a depth of 10 mm in a ICRU sphere, at given atmospheric altitude $h$ induced by a primary cosmic ray particle is determined as:

\begin{equation}
H^{*}(10) = \sum_{i} \int\limits_{T^{'}(\lambda_{m})}^{\infty} \int\limits_{\Omega} J_{i}(T{'}) Y_{i}^{*}(T^{'},h) d\Omega dT^{'}
\label{simp_eqn6}
   \end{equation}
\noindent where $Y_{i}^{*}(T^{'})$ is the ambient dose equivalent yield function determined in a way similar to equation (4). Here we adopt the data sets for an isotropic particle fluence from Appendix 2 of \cite{Pelliccioni2000} i.e. the fluence to ambient dose equivalent conversion coefficients for the described above secondaries. 

 \begin{figure} [!ht]
\centering \resizebox{\columnwidth}{!}{\includegraphics{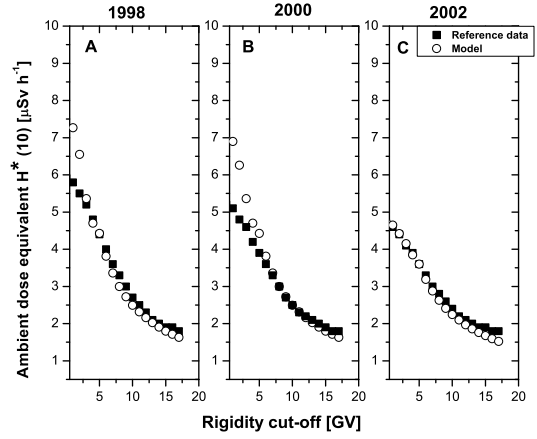}}
 \caption{Computed ambient dose equivalent $H^{*}(10)$ compared with reference data \cite{Menzel2010} at the altitude of 35 kft a.s.l. for different periods with various solar activity a) January 1998  b) January 2000 c) January 2002. \label{fig3}}
 \end{figure}

 \begin{figure} [!ht]
\centering \resizebox{\columnwidth}{!}{\includegraphics{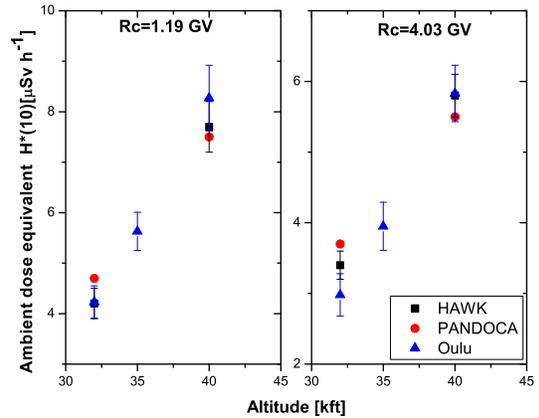}}
 \caption{Computed ambient dose equivalent $H^{*}(10)$ with Oulu model compared with PANDOCA model and recent measurements at two geographic points and at various altitudes as denoted in the legend  \label{fig4}}
 \end{figure} 

According to our computation the largest contribution to effective, accordingly ambient dose equivalent at near ground and ground altitudes is due to secondary muons and neutrons, while at aviation altitudes of about 10 km a.s.l. and at low cut-off rigidity, neutrons and protons together account for about 80$\%$ of the exposure. However, at high rigidity cut-off regions of about $R_{c}$ = 15 GV, the contribution of protons is smaller and the components of electromagnetic cascade take over.

The model is applied for computation of ambient dose equivalent $H^{*}(10)$ at altitude of 35 kft a.s.l. It is compared with reference data, shown in (Fig.3), which span periods corresponding to different solar activity, namely from solar minimum to solar maximum \cite{Menzel2010}. A good agreement is achieved, specifically in mid and high rigidity cut-off regions. The observed marginal difference for the period 1998 and 2000 in the region of low rigidity cut-off is due to modulation effects. In addition, a comparison with recent measurements (flight campaigns over Germany and Norway) as well as with PANDOCA model is performed and good agreement is achieved details given in Fig.4 \cite{Matthia14, Meier16}. The model also agrees with various measurements at different locations and latitudes  \cite{Lee15, Meier16b}

 \begin{figure} [!ht]
\centering \resizebox{\columnwidth}{!}{\includegraphics{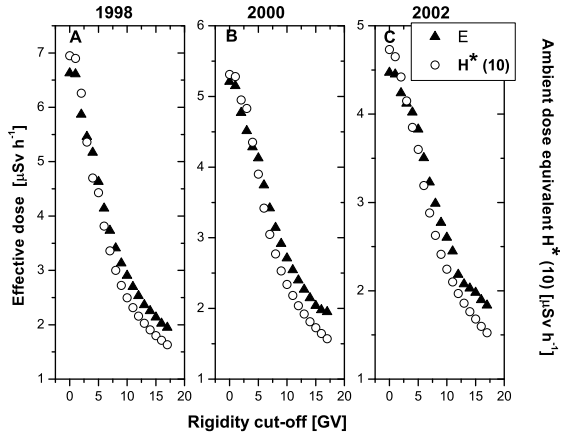}}
 \caption{Comparison of computed effective dose $E$ and ambient dose equivalent $H^{*}(10)$ at the altitude of 35 kft a.s.l. for different periods with various solar activity a) January 1998  b) January 2000 c) January 2002. \label{fig6}}
 \end{figure}

\begin{figure} [!ht]
\centering \resizebox{\columnwidth}{!}{\includegraphics{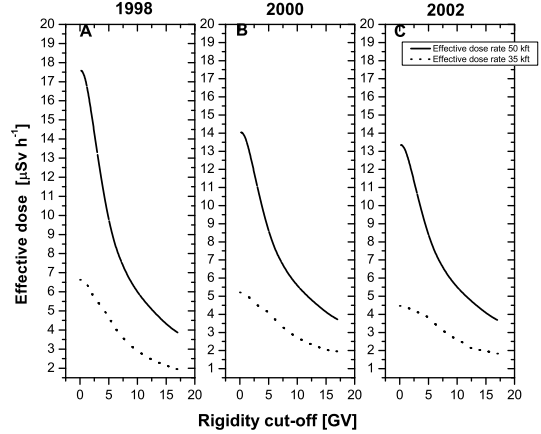}}
 \caption{Computed effective dose $E$ at the altitudes of 35 kft and 50 kft a.s.l. for different periods with various solar activity a) January 1998  b) January 2000 c) January 2002.  \label{fig7}}
 \end{figure} 

The estimated with the model effective dose $E$ at the altitudes of 35 kft is compared with the computed ambient dose equivalent $H^{*}(10)$ for different periods with various solar activity (Fig. 5). According to the presented in Fig.5 results the ambient dose equivalent $H^{*}(10)$ is a conservative approach for the exposure in low rigidity cut-off regions, while the effective dose is a conservative approach above some 5 GV. Even the ambient dose equivalent $H^{*}(10)$ is not a conservative estimate for cosmic radiation exposure at aviation altitudes, specifically in high rigidity regions, it is regarded as an acceptable approximation for effective dose. 

The model is applicable in a whole atmosphere. In Fig. 6 is resented an example of the computed effective dose $E$ at altitudes of 35 kft and 50 kft a.s.l. for different periods corresponding to  various solar activity level. There is a considerable increase of the computed effective dose rate as a function of the altitude a.s.l., specifically in a low rigidity cut-off region. In addition, the model was applied to estimate the contribution of cosmic radiation at high mountain altitude and good agreement with measurements is achieved, the details given elsewhere \cite{Mishev16d} . 

\section{APPLICATION OF THE MODEL FOR DOSE COMPUTATION DURING GLE EVENT}
During major GLEs, the dose rate is a superposition of the contribution of GCRs and SEPs, the latter possessing an essential anisotropic part, specifically during the event onset. As the first step we derive the SEP spectra and anisotropy outside the magnetosphere using data from the global NM network (Fig.1) and the method described in subsection II.A. In order to consider explicitly the anisotropy we compute the asymptotic cones in the region of interest in a grid of 5$^{\circ}$  $\times$ 5$^{\circ}$. As example we consider one of the strongest events of Solar Cycle 23, namely GLE 70. On 13 December 2006, NOAA active region 10930, located at S06W26, triggered a X3.4/4 B solar flare which reached maximum at 2:40 UT. It was associated with Type II and Type IV radio bursts and a fast full-halo CME accompanied by a strong solar proton event \cite{Moraal12, Gopalswamy2012}. The derived spectra and pitch angle distribution assuming modified power law rigidity spectrum and Gauss like angular distribution are presented in Fig.7, accordingly Fig.8, the details given elsewhere \cite{Mishev16}. During the event's onset, SEPs had a hard spectrum, strong anisotropy of a beam like SEP flux. The derived rigidity spectrum was gradually softening throughout the event (Fig.7). The steepening of the spectrum was moderate after the main phase of the event resulting in pure power law rigidity spectrum during the late phase of the event. Accordingly the derived anisotropy was high during the event's onset, but the SEPs angular distribution considerably broadened out during the event (Fig.8).

 \begin{figure} [!ht]
\centering \resizebox{\columnwidth}{!}{\includegraphics{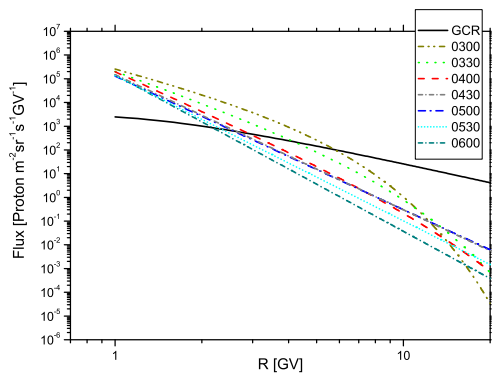}} 
 \caption{Derived rigidity spectra of SEPs during the GLE 70. \label{fig7}}
 \end{figure}

 \begin{figure} [!ht]
\centering \resizebox{\columnwidth}{!}{\includegraphics{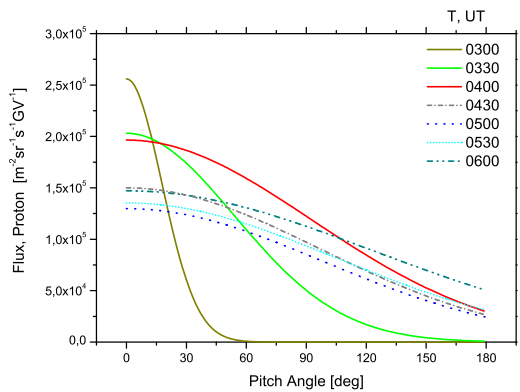}} 
 \caption{Derived pitch angle distributions of SEPs during the GLE 70. \label{fig8}}
 \end{figure} 
 
With the derived spectral and angular characteristics of SEPs during the GLE 70 and using the model described in section II.B we compute the effective dose rate at 35 kft during the event. During the initial phase of GLE 70 the computed effective dose was about 40-50 $\mu$Sv.h$^{-1}$ (Fig.9), while during the main (Fig.10) and late phases (Fig.11) the contribution of SEPs to the dose is comparable to the average due to GCR. All the computations are performed in a region with $R_{c}$ $\le$ 1 GV, where the expected exposure is maximal. The computed effective dose rate is in a very good agreement with previous estimations \citep{Matthia2009b}.

 \begin{figure} [!ht]
\centering \resizebox{\columnwidth}{!}{\includegraphics{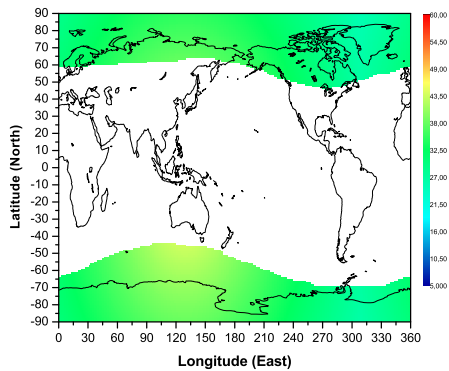}}
 \caption{Effective dose rate at the altitude of 35 kft a.s.l. during the initial phase of GLE 70 on 13 December 2006 in a region with $R_{c}$ $\le$ 1 GV. \label{fig9}}
 \end{figure}

\begin{figure} [!ht]
\centering \resizebox{\columnwidth}{!}{\includegraphics{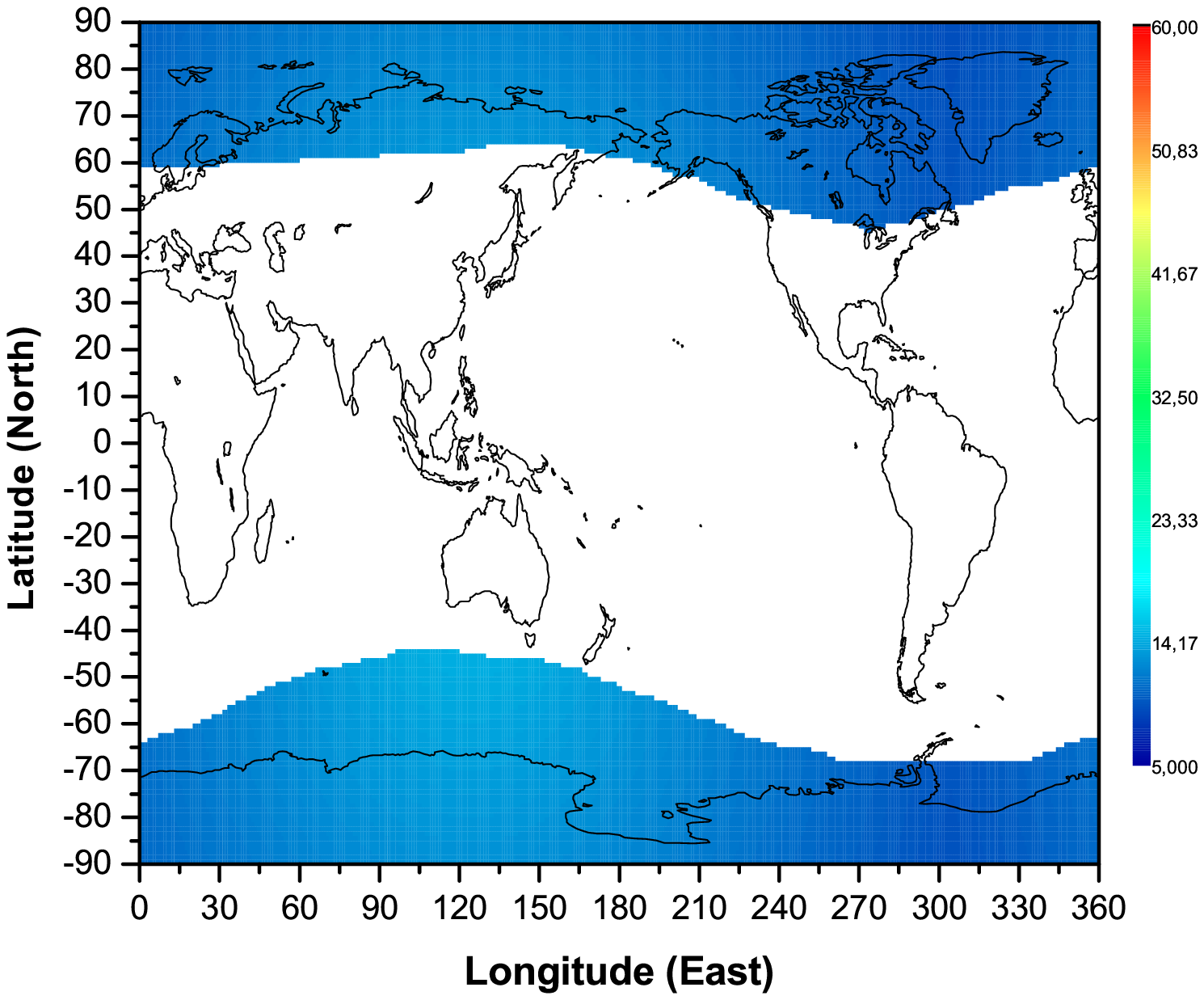}}
 \caption{Effective dose rate at the altitude of 35 kft a.s.l. during the main phase of GLE 70 on 13 December 2006 in a region with $R_{c}$ $\le$ 1 GV. \label{fig10}}
 \end{figure}

 \begin{figure} [!ht]
\centering \resizebox{\columnwidth}{!}{\includegraphics{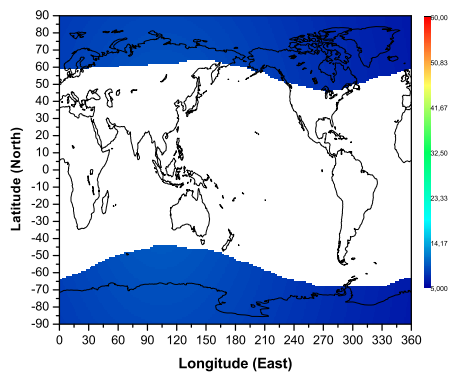}}
 \caption{Effective dose rate at the altitude of 35 kft a.s.l. during the late phase of GLE 70 on 13 December 2006 in a region with $R_{c}$ $\le$ 1 GV. \label{fig11}}
 \end{figure}

\section{Discussion and Conclusion}
According to the common definition, space weather refers to the dynamic, variable conditions on the Sun, solar wind and Earth's magnetosphere and ionosphere, that can compromise the performance and reliability of spacecraft and ground-based systems and can endanger human life or health \cite{Lilensten2009}. An important part of this field is the careful and precise assessment of aircrew exposure due to CR, specifically during GLEs. Therefore, the development and validation of models as well as the improvement of the existing models for dose assessment is of a big importance.

Here we present a new full chain numerical model for assessment of effective dose and/or ambient dose equivalent at flight altitudes. The model is based on NM data and newly computed yield functions for conversion of secondary particle fluence to effective and/or ambient dose equivalent. The full chain consists of several consecutive steps, namely derivation of SEP spectra and angular distribution by modelling NM responses, subsequent application of yield function(s), which  allow one to compute the exposure at different conditions (altitude, geographic position, solar activity level etc...). It was shown that the model demonstrates good agreement with reference data for the contribution of GCRs to the exposure at flight and high mountain altitudes. It is applied for computation of effective dose rate during a major GLE event, specifically in a low rigidity cut-off region. The dose rates are estimated in the course of the event, explicitly considering the primary particle flux variation over time. Therefore, the model is fully operational and could be used for space weather applications. Moreover, using a precomputed data-base of asymptotic trajectories of NMs (fast solution of the inverse problem i.e. derivation of SEP spectral and angular characteristics in a quasi-real time ) the model allows one to nowcast the exposure during major GLEs.

\bigskip % extra skip inserted
\begin{acknowledgments}
This work was supported by the Center of Excellence ReSoLVE (project No. 272157) of the Academy of Finland. We acknowledge all the colleagues from the neutron monitor stations, who kindly provided the data used for the analysis of GLE 70 included in GLE database, namely: Alma Ata, Apatity, Barentsburg, Calgary, Cape Schmidt, Forth Smith, Hermanus, Inuvik, Irkutsk, Jungfraujoch, Kerguelen, Kiel, Kingston, Lomnicky \v{S}tit, Magadan, Mawson, McMurdo, Moscow, Nain, Norilsk, Oulu, Peawanuck, Rome, Sanae, Terre Adelie, Thule, Tixie, Yakutsk. 
\end{acknowledgments}

\bigskip % extra skip inserted

\end{document}